\documentclass[reprint,
 amsmath,amssymb,
 aps,
]{revtex4-2}

\usepackage{graphicx}
\usepackage{dcolumn}
\usepackage{bm}
\usepackage{xcolor}

\begin{document}

\title{Space-time wave packets propagating a kilometer in air}

\author{Layton A. Hall$^{1}$}
\author{Miguel A. Romer$^{1}$}
\author{Bryan L. Turo$^{1}$}
\author{Tina M. Hayward$^{2}$}
\author{Rajesh Menon$^{2}$}
\author{Ayman F. Abouraddy$^{1,*}$}

\affiliation{$^{1}$CREOL, The College of Optics \& Photonics, University of Central~Florida, Orlando, FL 32816, USA}
\affiliation{$^{2}$Department of Electrical and Computer Engineering, University of Utah, Salt Lake City, Utah 84112, USA}
\affiliation{$^*$Corresponding author: raddy@creol.ucf.edu}

\begin{abstract}
We report on the diffraction-free propagation of space-time wave packets (STWPs) -- a class of propagation-invariant pulsed beams -- for $\sim\!1$~km in an open-air laser range in a low-turbulence scenario. Making use of $\approx\!100$-fs pulses (bandwidth $\sim\!25$~nm) at a wavelength of $\approx\!1$~$\mu$m, we construct an STWP with a transverse width of $\approx\!2$~mm that expands to $\approx\!3$~mm after $\sim\!500$~m, and another that expands from $\approx\!8$~mm to $\approx\!10$~mm after 1~km. The propagation of the STWPs is compared to Gaussian wave packets of the same transverse spatial width and bandwidth. We establish a theoretical model that accounts for the significant factors limiting the STWP propagation distance and suggests the path to further extending this distance.
\end{abstract}


\maketitle

Since the introduction of the Bessel beam in 1987 \cite{Durnin87PRL}, the potential utility of so-called `diffraction-free' monochromatic beams \cite{McGloin05CP,Turunen10PO} in long-distance propagation has been of continued interest, with potential applications in free-space communications. The experimental tests carried out to date demonstrated propagation distances typically on the order of meters \cite{Turunen88AO,Cox92JOSAA,Vetter19LPR,Stoyanov21OE}. Exceptions include 1-km experiments making use of a Bessel-like beam with controlled wavefront spherical aberrations \cite{Aruga99AO}, and most recently using an auto-focusing beam (a circularly symmetric Airy beam at a wavelength of 532~nm) whose initial 6-mm-diameter peak extends to 9~mm after 1~km \cite{Zhang19APLP}.

We have recently investigated a family of propagation-invariant (diffraction-free and dispersion-free) pulsed beams dubbed `space-time wave packets' (STWPs) \cite{Kondakci17NP,Yessenov22AOP}. STWPs are endowed with angular dispersion \cite{Torres10AOP}; i.e., each wavelength travels at a prescribed angle with respect to the propagation axis. In contrast, conventional tilted pulse fronts \cite{Fulop10Review} that are also endowed with angular dispersion (AD) are \textit{not} propagation invariant. We have recently uncovered that the AD undergirding STWPs is `non-differentiable'; i.e., the derivative of the propagation angle with respect to wavelength is not defined at a particular wavelength \cite{Hall21OL,Hall21OLnormalGVD,Hall22OEConsequences}. The non-differentiability of the underlying AD profile is the key to the unique attributes of STWPs, rendering them distinct from tilted pulse fronts. Besides propagation invariance, STWPs have tunable on-axis group velocity \cite{Salo01JOA,Efremidis17OL,Wong17ACSP2,Kondakci19NC} and group-velocity dispersion \cite{Yessenov21ACSP}, and exhibit self-healing \cite{Kondakci18OL} and anomalous refraction phenomena \cite{Bhaduri20NP}.

Initial demonstrations of STWPs verified propagation invariance over small distances, which are nevertheless significantly larger than the Rayleigh range of a Gaussian beam having the same initial spatial width. We have increased the propagation distance $L_{\mathrm{max}}$ of STWPs from $\approx\!100$~mm in \cite{Kondakci17NP}, to $\approx\!6$~m in a laboratory environment \cite{Bhaduri18OE}, and $\approx\!70$~m after directing the beam out of the laboratory and down a service chase in our research building \cite{Bhaduri19OL}. Our previous work predicted theoretically the possibility of extending $L_{\mathrm{max}}$ to the kilometer range.

Here we report on propagation measurements for STWPs over $\sim\!1$~km in the open environment of a laser range in Florida (TISTEF: Townes Institute Science and Technology Experimentation Facility). We first establish a theoretical model that accounts for the various factors that limit the propagation distance of an STWP. Based on this model we synthesize STWPs at a wavelength $\lambda_{\mathrm{o}}\!\sim\!1064$~nm and bandwidth $\Delta\lambda\!\sim\!25$~nm. The transverse width of one STWP expands from $\Delta x\approx\!2$~mm to $\Delta x\!\approx\!3$~mm after 500~m; a Gaussian wave packet expands over the same distance to $\approx\!400$~mm (Rayleigh range $z_{\mathrm{R}}\!\approx\!3$~m). The width of another STWP expands from $\Delta x\!\approx\!8$~mm to $\approx\!10$~mm after 1~km; a corresponding Gaussian wave packet expands to $\approx\!160$~mm ($z_{\mathrm{R}}\!\approx\!50$~m). Our model points to the improvements required to extend $L_{\mathrm{max}}$ to the 10-km range and beyond, which appears to be within current capabilities.

\begin{figure*}[t!]
\centering
\includegraphics[width=17.6cm]{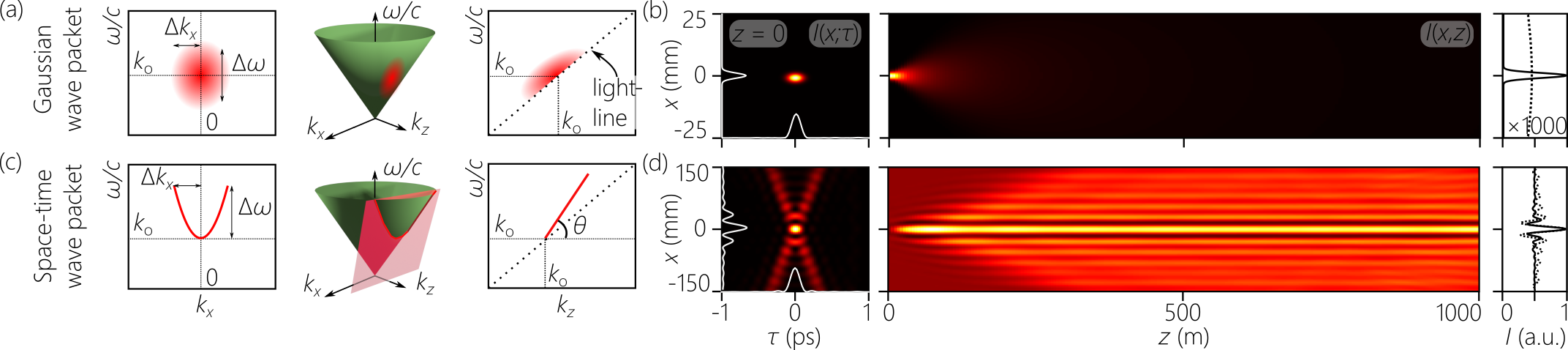}
\caption{(a) Representation of the spectral support for a separable pulsed Gaussian beam on the surface of the light-cone, along with spectral projections onto the $(k_{x},\tfrac{\omega}{c})$ and $(k_{z},\tfrac{\omega}{c})$ planes. (b) Spatio-temporal profile of the pulsed Gaussian beam $I(x;\tau)$ at $z\!=\!0$, and the time-averaged intensity $I(x,z)$. Here $\Delta k_{x}\!=\!0.7$~rad/mm, $\Delta x\!=\!2.5$~mm, $\lambda_{\mathrm{o}}\!=\!1064$~nm, $\Delta\lambda\!=\!25$~nm, $\Delta\tau\!=\!100$~fs.  The rightmost panel depicts $I(x,z)$ at $z\!=\!0$ (solid) and at $z\!=\!1000$~m (dashed). (c,d) Same as (a,b) for an STWP with $\theta\!=\!45.00001^{\circ}$, $\delta\lambda\!=\!0.5$~nm, and all other parameters are the same as for the pulsed Gaussian beam.}
\label{Fig:LightCones}
\end{figure*}

We compare conventional Gaussian wave packets (pulsed beams) in which the spatial and temporal degrees of freedom are separable, and STWPs in which they are not. We write the field in terms of a carrier and slowly varying envelope, $E(x,z;t)\!=\!e^{i(k_{\mathrm{o}}z-\omega_{\mathrm{o}}t)}\psi(x,z;t)$, where $\omega_{\mathrm{o}}$ is a carrier frequency, $k_{\mathrm{o}}\!=\!\omega_{\mathrm{o}}/c$, and $c$ is the speed of light in vacuum. We make use here of only one transverse spatial dimension $x$ and hold the field uniform along $y$. Recent developments have led to synthesizing STWPs modulated in both $x$ and $y$ \cite{Guo21Light,Pang22OE,Yessenov22NC}, and we will make use of them for long-distance experiments in the near future. The envelope of the Gaussian wave packet is:
\begin{equation}
\psi(x,z;t)\!=\!\iint\!dk_{x}d\Omega\widetilde{\psi}(k_{x},\Omega)e^{ik_{x}x}e^{i(k_{z}-k_{\mathrm{o}})z}e^{-i\Omega t},
\end{equation}
where $k_{x}$ and $k_{z}$ are the transverse and axial wave numbers, respectively, $\Omega\!=\!\omega-\omega_{\mathrm{o}}$, and the spatio-temporal spectrum $\widetilde{\psi}(k_{x},\Omega)$ is the 2D Fourier transform of $\psi(x,0;t)$. The spectral support of such a wave packet is a 2D domain on the surface of the light-cone associated with the free-space dispersion relationship $k_{x}^{2}+k_{z}^{2}\!=\!(\tfrac{\omega}{c})^{2}$, and the spectral projections onto the $(k_{x},\tfrac{\omega}{c})$ and $(k_{z},\tfrac{\omega}{c})$ are also 2D domains [Fig.~\ref{Fig:LightCones}(a)]. The spatio-temporal spectrum of conventional wave packets is typically separable $\widetilde{\psi}(k_{x},\Omega)\!\approx\!\widetilde{\psi}_{x}(k_{x})\widetilde{\psi}_{t}(\Omega)$, which is manifest in the initial spatio-temporal intensity profile $I(x,0;t)\!=\!|\psi(x,z\!=\!0;t)|^{2}$ [Fig.~\ref{Fig:LightCones}(b)]. In the narrowband, paraxial regime we have $k_{z}(k_{x},\omega)\!\approx\!\tfrac{\omega}{c}-\tfrac{k_{x}^{2}}{2k_{\mathrm{o}}}$, which leads to dephasing of the spatial frequencies along $z$, and thus diffraction of the time-averaged intensity $I(x,z)\!=\!\int\!dt\,I(x,z;t)$ [Fig.~\ref{Fig:LightCones}(b)].

In contrast, the spatio-temporal spectrum of STWPs is not separable, and ideally a one-to-one relationship between $k_{x}$ and $\Omega$ is enforced such that $\tfrac{\Omega}{\omega_{\mathrm{o}}}\!\approx\!\tfrac{k_{x}^{2}}{2k_{\mathrm{o}}^{2}(1-\cot{\theta})}$, and thus $\Omega\!=\!(k_{z}-k_{\mathrm{o}})\widetilde{v}$, which is equivalent to restricting the spectral support on the light-cone to a 1D trajectory at its intersection with a plane making an angle $\theta$ (the spectral tilt angle) with the $k_{z}$-axis [Fig.~\ref{Fig:LightCones}(c)]. The STWP envelope takes the form:
\begin{equation}
\psi(x,z;t)=\int\!d\Omega\widetilde{\psi}(\Omega)e^{i\{k_{x}x-\Omega(t-z/\widetilde{v})\}}=\psi(x,0;t-z/\widetilde{v}),
\end{equation}
which travels rigidly in free space at a group velocity $\widetilde{v}\!=\!c\tan{\theta}$ \cite{Kondakci19NC}. The spatio-temporal intensity profile $I(x,z;t)$ is X-shaped in any axial plane in this ideal limit [Fig.~\ref{Fig:LightCones}(d)].

\begin{figure}[t!]
\centering
\includegraphics[width=8.6cm]{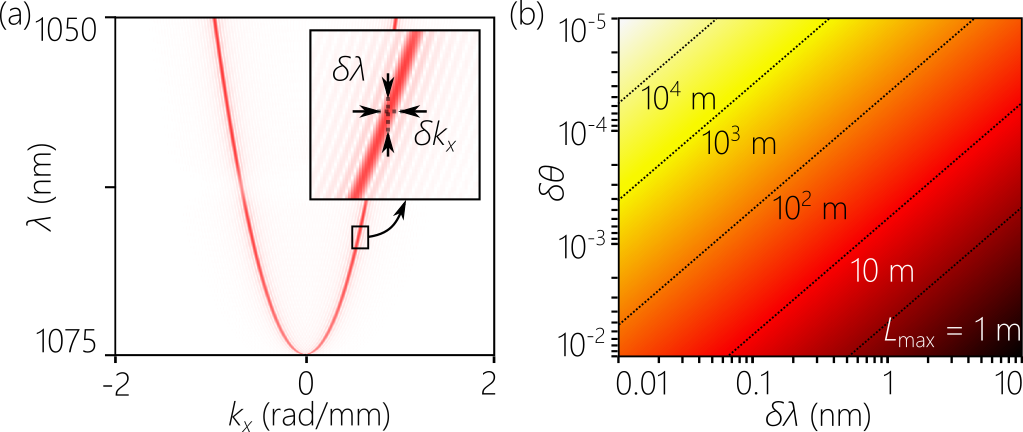}
\caption{(a) Calculated spatio-temporal spectrum of an STWP with $W_{\mathrm{A}}\!=\!150$~mm and $\delta\theta\!=\!10^{-5}$, highlighting the spectral uncertainty $\delta\lambda$ and spatial uncertainty $\delta k_{x}$ in the inset. (b) Calculated propagation distance $L_{\mathrm{max}}$ with the spectral uncertainty $\delta\lambda$ and offset in spectral tilt angle $\delta\theta\!=\!\theta-45^{\circ}$.}
\label{Fig:SpectralUncertainty}
\end{figure}

However, the ideal delta-function correlation between $k_{x}$ and $\Omega$ [Fig.~\ref{Fig:LightCones}(c)] cannot be attained in practice because it implies an infinite energy. Instead, a finite spectral uncertainty arises in the association between $k_{x}$ and $\Omega$ [Fig.~\ref{Fig:SpectralUncertainty}(a)], in which case $\widetilde{\psi}(k_{x},\Omega)\!\rightarrow\!\widetilde{\psi}(\Omega)\widetilde{g}(k_{x}-k_{x,\mathrm{o}}(\Omega))$, where $\widetilde{g}(\cdot)$ is a narrow function of width $\delta k_{x}$, and $k_{x,\mathrm{o}}(\Omega)$ is the spatial frequency associated with $\Omega$ in the ideal limit (in absence of spectral uncertainty). The spatial uncertainty $\delta k_{x}$ is associated with a spectral uncertainty $\delta\omega$ via $\tfrac{\delta\omega}{\omega_{\mathrm{o}}}\!\approx\!\tfrac{k_{x}\delta k_{x}}{k_{\mathrm{o}}^{2}|1-\cot{\theta}|}$ [Fig.~\ref{Fig:SpectralUncertainty}(a)]. Thus, rather than a mathematical parabola, the spatio-temporal spectrum projected onto the $(k_{x},\lambda)$-plane has a finite `thickness': eack spatial frequency $k_{x}$ is associated with a finite temporal bandwidth $\delta\lambda$. This spectral uncertainty is one of the two key parameters that determines the propagation distance $L_{\mathrm{max}}$ for an STWP, the other being $\theta$: $L_{\mathrm{max}}\!\sim\!\tfrac{c}{\delta\omega|1-\cot{\theta}|}$ \cite{Yessenov19OE}. To increase $L_{\mathrm{max}}$ one must reduce $\delta\omega$ and have $\theta\!\rightarrow\!45^{\circ}$; $\theta\!=\!45^{\circ}$ corresponds to a plane-wave pulse \cite{Yessenov19PRA}. Defining an offset in the spectral tilt angle $\delta\theta\!=\!\theta-45^{\circ}$, we have $L_{\mathrm{max}}\!\approx\!\tfrac{c}{2\delta\omega\delta\theta}$ for small $\delta\theta\!\ll\!1$ and $\delta\lambda\!<\!\Delta\lambda$ \cite{Bhaduri19OL}; see Fig.~\ref{Fig:SpectralUncertainty}(b).

\begin{figure}[t!]
\centering
\includegraphics[width=8.6cm]{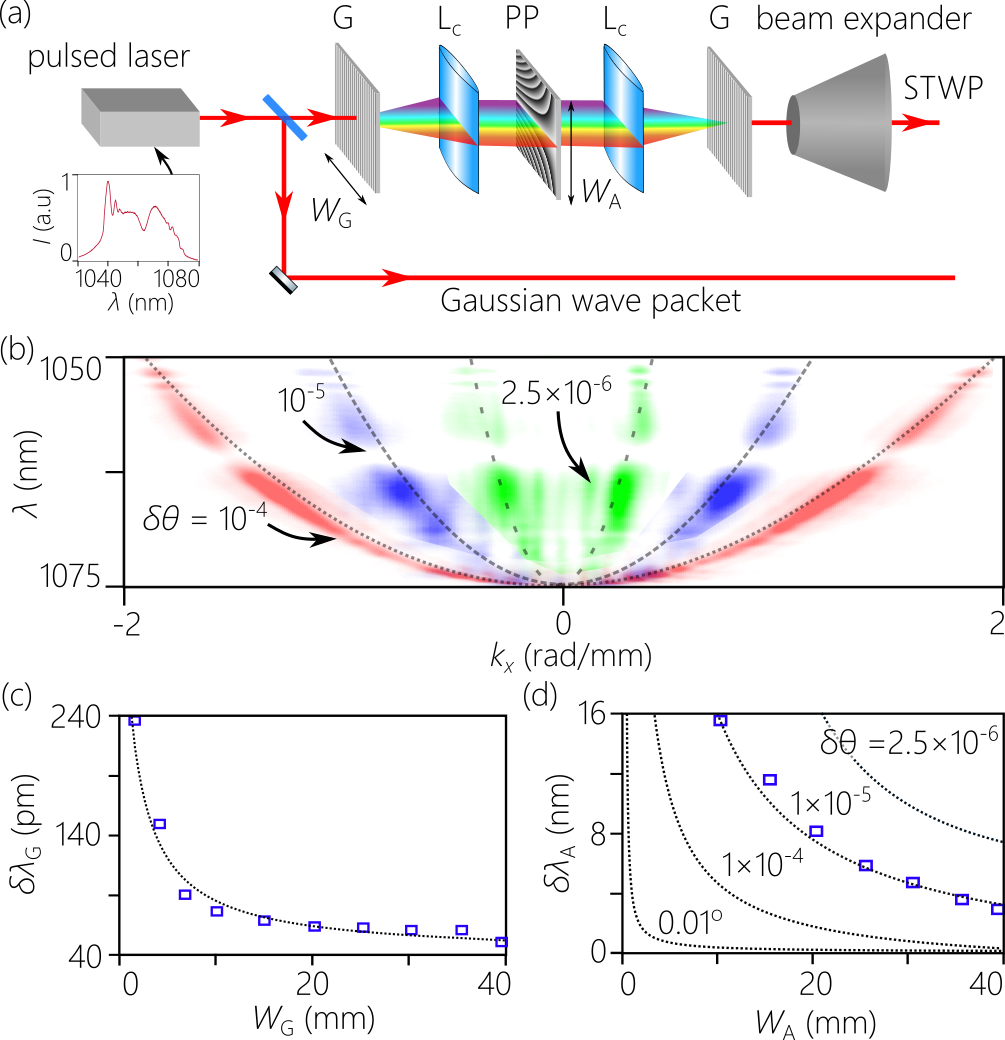}
\caption{(a) Schematic of the setup for synthesizing and launching STWPs. G: Diffraction grating; L$_{\mathrm{c}}$: cylindrical lens; PP: phase plate. The laser spectrum is shown in the inset. (b) Measured spatio-temporal spectra produced by phase plates corresponding to different $\delta\theta$. The dotted curves are theoretical expectations. (c) Measured spectral uncertainty $\delta\lambda$ produced by the grating width $W_{\mathrm{G}}$ and (b) the phase plate aperture $W_{\mathrm{A}}$.}
\label{Fig:Experiment}
\end{figure}

To reach large values of $L_{\mathrm{max}}$, we must first identify the experimental factors that limit $\delta\theta$ and $\delta\lambda$. The experimental arrangement we use \cite{Kondakci18OE} is depicted in Fig.~\ref{Fig:Experiment}(a). We utilize pulses from a mode-locked laser (Spark Laser, Alcor) of width $\Delta T\!\approx\!100$~fs, bandwidth $\Delta\lambda\!\approx\!25$~nm, central wavelength $\lambda_{\mathrm{o}}\!\approx\!1.064$~$\mu$m, and average power $\approx\!2$~W. The initial $\approx\!2$-mm beam width is magnified with a $20\times$ beam expander and directed to a diffraction grating (1200~lines/mm, $50\times50$~mm$^2$). The -1~diffraction order is collimated by a cylindrical lens (focal length $f\!=\!250$~mm), and the spatially resolved spectrum traverses a transparent phase plate that modulates the spatial phase to associate a prescribed pair of symmetric spatial frequencies $\pm k_{x}(\lambda)$ to each wavelength $\lambda$, before the pulse is reconstituted by an identical grating-lens pair. The phase plate is fabricated via gray-scale lithography using the procedure outlined in \cite{Mohammad17SR}. The phase plate also acts as a limiting aperture in the system of width $W_{\mathrm{A}}$ (here $W_{\mathrm{A}}\!=\!40$~mm).

In considering the limit on $\delta\theta$, it is important to note that $\theta$ is \textit{not} an angle in physical space, but is an angular parameter that determines the rate of change of $k_{x}$ across the bandwidth $\Delta\lambda$. The spatio-temporal spectrum is captured by placing a mirror after the second cylindrical lens to deflect the field beam before the second grating, and focus the spatio-temporal spectrum using a spherical lens in a $2f$ configuration to a CCD camera. The measurements in Fig.~\ref{Fig:Experiment}(b) indicate that $\delta\theta$ can be made very small. The results in as shown in Fig.~\ref{Fig:Experiment}(b) correspond to $\delta\theta\!\approx\!10^{-4}$, $10^{-5}$, and $2.5\times10^{-6}$ (all in degrees); the latter two of which are utilized in our long-distance experiments. Therefore, the span of values of $\delta\theta$ in Fig.~\ref{Fig:SpectralUncertainty}(b) is currently accessible.

With regards to the second factor limiting $L_{\mathrm{max}}$, the spectral uncertainty $\delta\lambda$, we previously identified the spectral resolution of the grating $\delta\lambda_{\mathrm{G}}\!=\!\tfrac{\lambda_{\mathrm{o}}}{n_{\mathrm{G}}W_{\mathrm{G}}}$, where $n_{\mathrm{G}}$ is the grating groove density and $W_{\mathrm{G}}$ is its width, as the main contributor, $\delta\lambda\!\approx\!\delta\lambda_{\mathrm{G}}$ \cite{Yessenov19OE,Kondakci19OL}. In Fig.~\ref{Fig:Experiment}(c) we plot the measured $\delta\lambda$ while varying the width $W_{\mathrm{G}}$ of the grating aperture [Fig.~\ref{Fig:Experiment}(a)]. We measure $\delta\lambda$ at the focal plane where the phase plate is placed by capturing the field with a single mode fiber coupled to an optical spectrum analyzer (Advantest AQ6317B), and the measurements plotted in Fig.~\ref{Fig:Experiment}(c) confirm the expected $\tfrac{1}{W_{\mathrm{G}}}$ dependence. Crucially, the contribution of $W_{\mathrm{G}}$ to $\delta\lambda$ is independent of $\theta$.

\begin{figure}[t!]
\centering
\includegraphics[width=8.6cm]{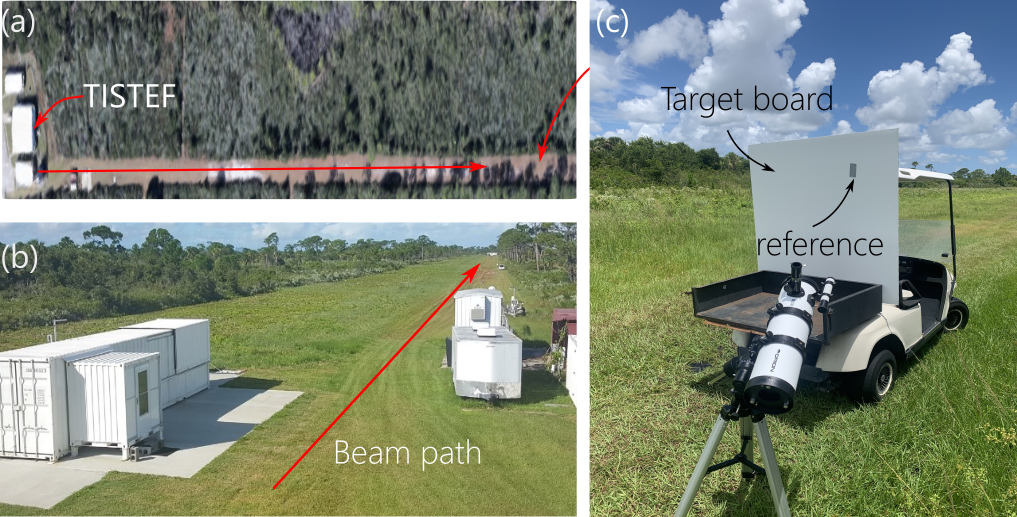}
\caption{(a) The TISTEF facility where the setup in Fig.~\ref{Fig:Experiment}(a) is placed. (b) The laser range along which the measurements performed. (c) Detection scheme on a target board used to image the STWP down the range.}
\label{Fig:TISTEF}
\end{figure}

The measurements shown in Fig.~\ref{Fig:Experiment}(c) indicate that the small values of $\delta\lambda$ needed to reach the km-range according to Fig.~\ref{Fig:SpectralUncertainty}(b) are accessible. However, in the course of our experiments we have identified a second factor contributing to $\delta\lambda$: the height of the phase plate $W_{\mathrm{A}}$ [Fig.~\ref{Fig:Experiment}(a)], which represents a limiting spatial aperture for the synthesis system. This aperture results in a new contribution to the spectral uncertainty we denote $\delta\lambda_{\mathrm{A}}$. it can be shown that $\delta\omega_{\mathrm{A}}\!\approx\!\tfrac{\pi c\lambda_{\mathrm{o}}}{W_{\mathrm{A}}\Delta x\delta\theta}$, and the total spectral uncertainty is $\delta\omega\!\approx\!\sqrt{(\delta\omega_{\mathrm{A}})^{2}+(\delta\omega_{\mathrm{G}})^{2}}$, which is exact if the spectral uncertainties are modelled with Gaussian spectral profiles. Because $\delta\lambda_{\mathrm{A}}$ depends on $\theta$, it must be measured after the phase plate. We carry out a spatial Fourier transform of the STWP along $x$ after the second cylindrical lens, and a single-mode fiber coupled to the optical spectrum analyzer captures the spectral width $\delta\lambda_{\mathrm{A}}$. From these results, it is clear that $W_{\mathrm{A}}$ needs to be enlarged to reach the kilometer scale.

\begin{figure*}[t!]
\centering
\includegraphics[width=17.6cm]{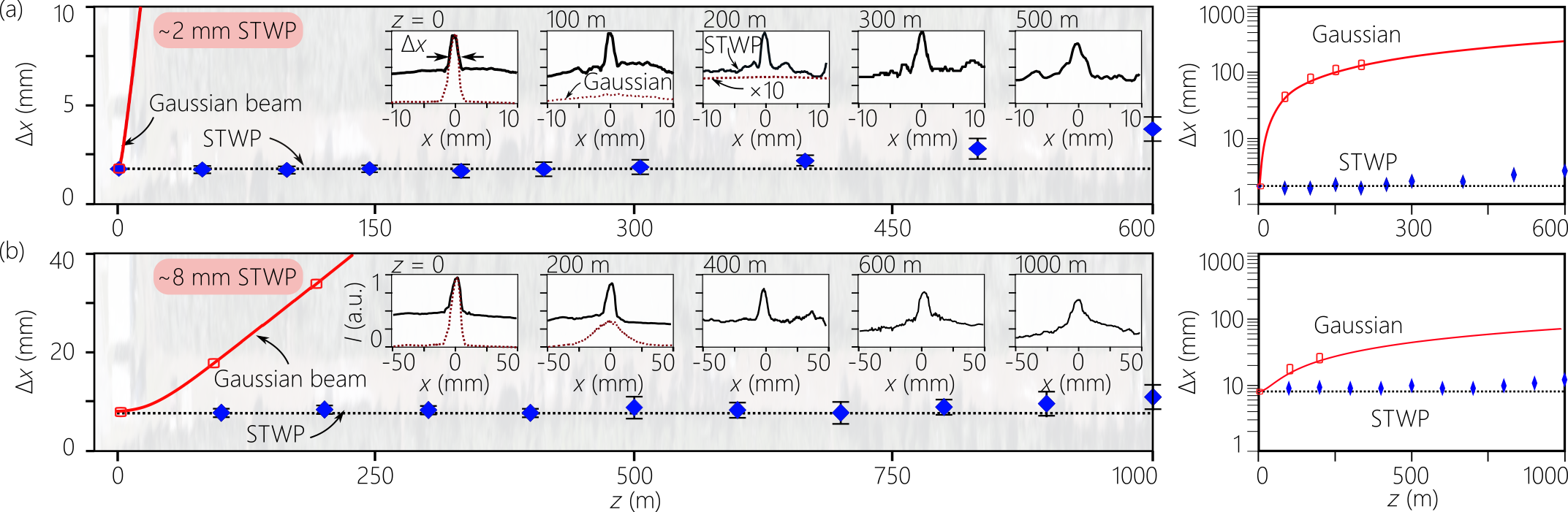}
\caption{Measured width $\Delta x$ along $z$ for an STWP and Gaussian wave packet that start with the same width: (a) $\Delta x\!\approx\!2$~mm and (b) $\Delta x\!\approx\!8$~mm. Insets are 1D sections through the wave packets along $z$. The right panels are the same results plotted with a vertical logarithmic scale to capture the diffractive spreading of the Gaussian wave packet. The solid curves are theoretical expectations.}
\label{Fig:KilometerData}
\end{figure*}

The TISTEF laser range facility in which we carried out our experiments is shown in Fig.~\ref{Fig:TISTEF}, which comprises a 1-km terrestrial laser range in the vicinity of NASA's Kennedy space center. Crucially for our measurements, TISTEF is capable of providing real-time atmospheric turbulence measurements. The transverse width of the STWP produced by the setup in Fig.~\ref{Fig:Experiment}(a) is enlarged from $\sim\!40$~mm to $\sim\!150$~mm using a combination of a negative lens ($f\!=\!-150$~mm, 50-mm diameter) and a positive lens ($f\!=\!600$~mm, 150-mm diameter), which provide a $4\times$ magnification and minimize the spherical aberrations to 1.2~wavelengths. The divergence in this system is $\approx\!0.03$~mrad, whereas the ideal divergence is $\approx\!0.01$~mrad as determined by Zemax calculations. The STWP is then launched from the laboratory through an open window onto the laser-range grass field [Fig.~\ref{Fig:TISTEF}(a)]. To detect the beam, we intercept it with a painted metal board on the back of a vehicle that moves to targeted distances down the range to $\approx\!1$~km from the laboratory. To locate the STWP as the vehicle moves down the range, we image the STWP on the board via a telescope to a CCD (equipped with a 20-nm-bandwidth spectral filter centered at 1064~nm) placed in the laboratory next to the synthesis setup [Fig.~\ref{Fig:TISTEF}(a)]. A second image is taken \textit{in situ} down the range next to the board with a smaller telescope to a similar CCD and spectral filter [Fig.~\ref{Fig:TISTEF}(b)]. In addition to the STWPs, we also launch down the range Gaussian wave packets with separable spatial and temporal degrees of freedom for comparison. These Gaussian wave packets are taken from the laser directly [Fig.~\ref{Fig:Experiment}(a)], so they have the same bandwidth $\Delta\lambda$ as the STWPs, and their transverse width is made to match $\Delta x$ for the STWPs.

The measurements taken down the laser range are plotted in Fig.~\ref{Fig:KilometerData} for beams of width $\Delta x\!\approx\!2$~mm [Fig.~\ref{Fig:KilometerData}(a)] and $\Delta x\!\approx\!8$~mm [Fig.~\ref{Fig:KilometerData}(b)]. In each case, we plot the measured $\Delta x$ along $z$ for an STWP and a reference Gaussian wave packet of equal initial width $\Delta x$ at $z\!=\!0$. It is clear in both cases that the width of the Gaussian wave packet increases rapidly with $z$, whereas that for the STWP is maintained for a significantly extended distance. In Fig.~\ref{Fig:KilometerData}(a), the width for the STWP increases from $\approx\!2$~mm to $\approx\!3$~mm at 500~m, whereas the width of the corresponding Gaussian wave packet increases to $\approx\!130$~mm at 200~mm (sunlight prevented measurements at longer distances). Extending the propagation distance of the Gaussian wave packet to 500~m, its width is $\sim\!120\times$ that of the STWP. In Fig.~\ref{Fig:KilometerData}(b), the width for the STWP increases from $\approx\!8$~mm to $\approx\!10$~mm at 1~km, whereas that for the corresponding Gaussian wave packet increases to $\approx\!36$~mm at 200~mm. At 1~km, the width of the Gaussian wave packet is $\sim\!20\times$ that of the STWP.

These results were obtained in the early morning with measured refractive-index structure parameter $C_{n}^{2}\!\approx\!10^{-14}$~m$^{-2/3}$; as such, the impact of turbulence was minimal. We have carried out measurements corresponding to $C_{n}^{2}\!\sim\!10^{-15}\!-\!10^{-13}$~m$^{-2/3}$, and have found that the root-mean-square error of the STWPs is consistently lower than those of the Gaussian wave packets. Future measurements will aim at achieving longer $L_{\mathrm{max}}$ according to the predictions in Fig.~\ref{Fig:SpectralUncertainty}(b) by further reducing $\delta\lambda$ (by increasing the aperture $W_{\mathrm{A}}$). Moreover, we will make use of STWPs that are localized in both transverse dimensions \cite{Yessenov22NC}. The transverse intensity profile in this case is circularly symmetric with a $\tfrac{1}{r}$ radial decay rate ($r$ is the radial coordinate) and no pedestal. Finally, note that the wavelength used here is $\sim\!1$~$\mu$m compared to $\sim\!0.5$~$\mu$m in \cite{Zhang19APLP}, and thus the 1-km propagation distance recorded here is equivalent to twice that in \cite{Zhang19APLP}.

In conclusion, we have demonstrated kilometer-scale propagation of STWPs in air and compared their diffractive spreading to that of Gaussian wave packets having the same spatial and temporal bandwidths as the STWPs. We observe minimal increase in the STWP transverse width with respect to that of the Gaussian wave packets. Our theoretical model for the propagation distance as determined by the experimental parameters indicates that longer distances ($\sim\!10$~km) are within reach. 

\section*{Funding}
U.S. Office of Naval Research (ONR) N00014-19-1-2192.
\section*{Acknowledgments}
We thank the TISTEF staff for assistance.
\section*{Disclosures}
The authors declare no conflicts of interest.

\section*{Data availability}
Data underlying the results presented in this paper are not publicly available at this time but may be obtained from the authors upon reasonable request.

\bibliography{diffraction}

\end{document}